# A Novel Coupled bES-FEM Formulation with SUPG stabilization for Thermo-Hydro-Mechanical Analysis in Saturated Porous Media

Zi-Qi TANG, PhD Student

State Key Laboratory of Intelligent Geotechnics and Tunnelling, Shenzhen University, Shenzhen 518060, Guangdong, China

College of Civil and Transportation Engineering, Shenzhen University, Shenzhen 518060, Guangdong, China

Department of Civil and Environmental Engineering, The Hong Kong Polytechnic University, Hung Hom, Kowloon, Hong Kong, China; Email: ziqi-scut.tang@connect.polyu.hk

Xi-Wen ZHOU, PhD Student

Department of Civil and Environmental Engineering, The Hong Kong Polytechnic University, Hung Hom, Kowloon, Hong Kong, China; Email: zhouxw26@mail2.sysu.edu.cn

Yin-Fu JIN, Professor (corresponding author)

State Key Laboratory of Intelligent Geotechnics and Tunnelling, Shenzhen University, Shenzhen 518060, Guangdong, China

College of Civil and Transportation Engineering, Shenzhen University, Shenzhen 518060, Guangdong, China

Email: yinfujin@szu.edu.cn

Zhen-Yu YIN, Professor (corresponding author)

Department of Civil and Environmental Engineering, The Hong Kong Polytechnic University, Hung Hom, Kowloon, Hong Kong, China; Email: zhenyu.yin@polyu.edu.hk

Qi ZHANG, Research Assistant Professor

Department of Civil and Environmental Engineering, The Hong Kong Polytechnic University, Hung Hom, Kowloon, Hong Kong, China; Email: q7zhang@polyu.edu.hk

**Abstract:**

Two primary types of numerical instabilities often occur in low-order finite element method (FEM) analyses of thermo-hydro-mechanical (THM) phenomena: (1) pressure oscillations arising improper interpolation of pressure and displacement fields; and (2) spatial oscillations induced by nonlinear convection terms in convection-dominated scenarios. In response to these issues, this paper proposes a novel stabilized edge-based smoothed FEM with a bubble function (bES-FEM) for THM analysis within saturated porous media. In the proposed framework, a cubic bubble function is first incorporated into ES-FEM to efficiently mitigate pressure oscillations that breach the Inf-Sup condition, and then the Streamline Upwind Petrov-Galerkin (SUPG) scheme is adopted in bES-FEM to effectively reduce the spurious oscillations in convection-dominated heat transfer scenarios. The accuracy of the bES-FEM with SUPG formulation for THM coupled problems is validated through a series of five benchmark tests. Moreover, the simulations of open-loop ground source energy systems demonstrate the proposed method's exceptional capability in tackling complex THM challenges in real-world applications. All the obtained results showcase the superiority of proposed bES-FEM with SUPG in eliminating the spatial and pressure oscillations, marking it as a promising tool for the exploration of coupled THM issues.

# 1. Introduction

Thermo-hydro-mechanical (THM) coupling has always been an important concern in geotechnical engineering (Cekerevac and Laloui, 2004; Cui et al., 2016; Vaziri, 1996), given its pervasive influence on many geotechnical problems, where the interplay among thermal, hydraulic, and mechanical behaviour within solids becomes critical. Typically, the pore water pressure in the saturated soil surrounding a structure rises with the increase of temperature (Abuel-Naga et al., 2007; Gens, 2010), potentially leading to thermal failure of soils. Additionally, the investigation of efficiency considerations in geothermal systems for building heating (Brandl, 2006) has attracted numerous research works.

Numerous experiments (Campanella and Mitchell, 1968; Gens et al., 2011; Lima et al., 2010; Mohajerani et al., 2012) have been conducted to investigate the temporal progression of excess pore fluid pressure and temperature in soils. In recent years, numerical tools (Abed and Sołowski, 2017; Cui et al., 2018; François et al., 2009; Zdravković, 1999) have been increasingly developed to simulate the THM coupled challenges, serving as a complement to experiments owing to their convenience and cost-effectiveness. Yu et al. (2024) proposed a single-point multi-phases semi-implicit stable stabilized material point method (MPM) to study THM problems in saturated porous media. Based on the smoothed particle hydrodynamics (SPH) algorithm, Mu et al. (2024) reported a total Lagrangian form of SPH to investigate the propagation of crack in rocks. Cui et al. (2018) developed the finite element method (FEM) software ICFEP for THM coupled problems of saturated soils and implemented it into geotechnical engineering analysis. Ni et al. (2023) introduced a peridynamics method into the FEM framework to describe the deformation of solid phases and crack propagation. Other numerical methods for THM coupling analysis in porous media are summarized in McClure and Horne (2014) and White and Phillips (2015). Although numerous numerical methods have been developed for THM coupling problems, there are still some limitations and challenges in saturated porous media. Among these challenges, two types of numerical oscillation problems warrant special attention. The first one is the pressure oscillation resulting from the inappropriate interpolation for pressure and displacement field that fails to satisfy the inf-sup condition, and the second one is spatial oscillation caused by the nonlinear convective term in the heat transfer process.

By employing the gradient smoothing techniques for linear elements, the smoothed finite element method (S-FEM) (Zeng and Liu, 2018) offers an alternative approach to address instabilities arising from inappropriate interpolation for pressure and displacement field. This method fundamentally relies on the concepts of using displacement and strain fields coupled with a smoothing operation (Chen et al., 2001). The smoothing domain customarily overlaps parts of multiple adjacent elements and typically involves a greater number of supporting nodes than a linear element (Nguyen-Xuan and Liu, 2013). Despite a greater computational bandwidth and cost necessitated by a smoothing domain with more nodes, this strategy yields "softer" solutions (Jiang et al., 2017; Liu and Trung, 2016; Nguyen-Xuan and Liu, 2013) that more readily satisfy the inf-sup conditions, effectively mitigating the over-stiffness problem encountered in linear FE analysis. This is because the propagation of non-local information within the smoothing domain enables S-FEM to consistently deliver solutions of higher accuracy compared to conventional FEM (Nguyen-Xuan and Liu, 2013; Zeng and Liu, 2018). In pursuit of desired softening effects, a suite of methodologies has emerged through distinct approaches to smoothing domain formulation. These include cell-based S-FEM (CS-FEM) (Liu et al., 2007), node-based S-FEM (NS-FEM) (Liu et al., 2009b; Zhang et al., 2022) and edge-based S-FEM (ES-FEM) (Liu et al., 2009a). Among these methods, ES-FEM stands out for its superior computational performance (Nguyen-Xuan and Liu, 2013). Although employing a smoothing domain alleviates the overly-stiffness problem in ES-FEM to an extent, it does not fully satisfy the inf-sup condition (Brezzi and Bathe, 1990; Liu et al., 2009a; White and Borja, 2008). However, this indicates that ES-FEM is not entirely free from volumetric locking, necessitating further research. To address this, we develop a bES-FEM formulation that introduces bubble functions into ES-FEM, resulting in a further softening effect and fully satisfying the inf-sup condition.

As for the spatial oscillation caused by the nonlinear convective term in the THM coupling problems, the instability is due to the standard Galerkin FEM framework (Brooks and Hughes, 1982) treating upstream and downstream nodes with equal weight, which contradicts the fact that upstream nodes should be more heavily weighted physically. In an attempt to eliminate this spurious oscillation, many stabilization methods have been developed including the Taylor-Galerkin method (Donea, 1984; Hawken et al., 1990), least-squares finite element method (Jiang, 1998; Kim and Reddy, 2021),

Variational MultiScale method (Ahmed et al., 2017; Ahmed and Rubino, 2018), Streamline Upwind Petrov-Galerkin (SUPG) (Brooks and Hughes, 1982; Wervaecke et al., 2012), and so on. To mitigate the spatial oscillations potentially induced by heat transfer in high-velocity flows (Cui and Wong, 2022), we adopt the SUPG scheme in this work and incorporate it into the bES-FEM formulation for the saturated soil analysis of THM coupling. Essentially, the merit of bES-FEM formulation with SUPG stabilization lies in the integration of six significant features: (i) the proficiency in simulating near-incompressible or low-permeability media, notably in overcoming excess pore water pressure instabilities. (ii) the ability to eliminate numerical oscillations in convection-dominated situations. (iii) the seamless integration approach that readily dovetails with existing FEM packages without additional degrees of freedom; (iv) the superiority to produce more accurate solutions compared to linear triangular elements; (v) the insensitivity to mesh distortion due to potential large deformations or other factors; (vi) the versatility in simulating single-phase phenomena, arbitrary two-phase coupled phenomena, or three-phase coupled phenomena within the current formulation.

This paper is organized as follows. Section 2 provides an overview of the THM coupled models for saturated soils, including the mass conservation, momentum conservation and energy conservation equations. Section 3 details the formulation of bES-FEM and the integrated SUPG scheme. The effectiveness of the current bES-FEM formulation with SUPG stabilization is validated by a series of benchmark tests in Section 4, ranging from simple to intricate cases. In Section 5, two scenarios, including the thermo-elastic loading consolidation and open-loop energy system, are investigated to illustrate the superiority of the bES-FEM formulation with SUPG stabilization. Conclusions are presented in Section 6.

## 2. Fundamental of thermo-hydro-mechanical coupled model

*2.1 Mass conservation equation*

For saturated soils, it is considered that a representative volume element (RVE) with volume $V$ consists of solid and fluid phases, which occupy volumes $V_s$ and $V_f$, respectively. The volume fraction of solid and pore fluid can be determined as

$$\phi_s = V_s / V, \quad \phi_f = V_f / V . \tag{1}$$

The closure condition of volume fraction requires $\phi_s + \phi_f = 1$ and sometimes $\phi_f$ is abbreviated as $\phi$, which is the so-called porosity. The partial densities of solid and pore fluid are thus defined as

$$\rho^s = \phi_s \rho_s, \quad \rho^f = \phi_f \rho_f, \tag{2}$$

where $\rho_s$ and $\rho_f$ are the intrinsic densities of solid and pore fluid, respectively. The total density of the mixture is $\rho = \rho^s + \rho^f$. Assuming no mass exchanges between the solid and fluid phases, the corresponding mass conservation equations (Zhang and Borja, 2021) can be stated as

$$\phi_s \frac{d\rho_s}{dt} + \rho_s \frac{d\phi_s}{dt} + \phi_s \rho_s \nabla \cdot \mathbf{v}_s = 0, \tag{3}$$

$$\phi_f \frac{d^f \rho_f}{dt} + \rho_f \frac{d^f \phi_f}{dt} + \phi_f \rho_f \nabla \cdot \mathbf{v}_f = Q_f, \tag{4}$$

in which $\mathbf{v}_s$ and $\mathbf{v}_f$ are the intrinsic velocities of solid and pore fluid, respectively. $Q_f$ is the pore fluid source. According to Lei et al. (2021) and Pinyol et al. (2018), the intrinsic densities of solid and fluid can be assumed to have the following exponential functions form:

$$\rho_s = \rho_s^0 \exp\left[-3\alpha_s (T - T_0)\right], \tag{5}$$

$$\rho_f = \rho_f^0 \exp\left[\frac{1}{K}(p - p_0) - 3\alpha_f (T - T_0)\right], \tag{6}$$

in which, $T$ and $p$ denote the mixture temperature and pressure. $\rho_s^0$ and $\rho_f^0$ are the densities of solid and fluid at the reference condition, i.e., $p = p_0$ and $T = T_0$. $K$ is the bulk modulus of the fluid. $\alpha_s$ is the linear thermal expansion coefficient of solid, and $\alpha_f$ is the linear thermal expansion coefficient of fluid.

Given that the solid's intrinsic density exhibits only minor variations within the RVE (Bui and Nguyen, 2017), substituting Eq.(5) into Eq.(3) yields the following equations:

$$\frac{d\phi_s}{dt} + \phi_s \nabla \cdot \mathbf{v}_s - 3\alpha_s \phi_s \frac{dT}{dt} = 0. \tag{7}$$

Assuming the pore fluid is incompressible (Yerro Colom, 2015), i.e., $K = +\infty$, a combination of Eq.(4) and Eq.(6) yields,

$$\frac{d^f \phi_f}{dt} + \phi_f \nabla \cdot \mathbf{v}_f - 3\alpha_f \phi_f \frac{d^f T}{dt} = \frac{Q_f}{\rho_f} \ . \tag{8}$$

Given that the material time derivative of the fluid phase relative to the solid phase is defined by

$\frac{d^f (\cdot)}{dt} = \frac{d^s (\cdot)}{dt} + \nabla(\cdot) \cdot (\mathbf{v}_f - \mathbf{v}_s)$, Eq.(8) can be further rewritten using Eq.(7) as

$$\begin{aligned}
&\frac{d^f \phi_f}{dt} + \phi_f \nabla \cdot \mathbf{v}_f - 3\alpha_f \phi_f \frac{dT}{dt} \\
&= \frac{d^s \phi_f}{dt} + \nabla \phi_f \cdot (\mathbf{v}_f - \mathbf{v}_s) + \phi_f \nabla \cdot \mathbf{v}_f - 3\alpha_f \phi_f \frac{dT}{dt} \\
&= \phi_s \nabla \cdot \mathbf{v}_s - 3\alpha_s \phi_s \frac{dT}{dt} + \phi_f \nabla \cdot \mathbf{v}_f + \nabla \phi_f \cdot (\mathbf{v}_f - \mathbf{v}_s) - 3\alpha_f \phi_f \frac{dT}{dt} \\
&= \phi_s \nabla \cdot \mathbf{v}_s - 3\alpha_s \phi_s \frac{dT}{dt} + \phi_f \nabla \cdot \mathbf{v}_s + \phi_f \nabla \cdot (\mathbf{v}_f - \mathbf{v}_s) + \nabla \phi_f \cdot (\mathbf{v}_f - \mathbf{v}_s) - 3\alpha_f \phi_f \frac{dT}{dt} \\
&= \nabla \cdot \mathbf{v}_s + \nabla \cdot (\phi_f \cdot (\mathbf{v}_f - \mathbf{v}_s)) - 3(\alpha_s \phi_s + \alpha_f \phi_f) \frac{dT}{dt} = \nabla \cdot \mathbf{v}_s + \nabla \cdot \mathbf{v}_r - 3\alpha \frac{dT}{dt} = \frac{Q_f}{\rho_f} \ ,
\end{aligned} \tag{9}$$

where $\mathbf{v}_r = \phi_f \cdot (\mathbf{v}_f - \mathbf{v}_s)$ is the Darcy velocity. Since the deformation of the solid corresponds to the overall deformation of the RVE, which includes both mechanical and thermal deformation, Eq.(9) is reformulated as

$$\nabla \cdot \mathbf{v} - 3\alpha \frac{dT}{dt} + \nabla \cdot \mathbf{v}_r = \frac{Q_f}{\rho_f} \tag{10}$$

in which $\mathbf{v}$ and $T$ are the solid skeleton velocity vector and temperature, respectively. $\alpha = \alpha_s \phi_s + \alpha_f \phi_f$ is the linear thermal expansion coefficient of the mixture. Notably, we assume an instantaneous thermal equilibrium between the soil and pore fluid, in other words, the temperatures of both the solid and fluid phases are equal to the RVE's temperature $T$.

*2.2 Momentum conservation equation*

Employing the Lagrangian description, the motion equation for solid skeleton is given as

$$\nabla \cdot \boldsymbol{\sigma} + \mathbf{b} = \mathbf{0} \ , \tag{11}$$

where $\boldsymbol{\sigma}$, $\mathbf{b}$, and $\mathbf{u}$ are the stress tensor, body force vector and nodal displacement of mixture, respectively. According to the effective stress principle, the relationship between the total stress $\boldsymbol{\sigma}$ in RVE and effective stress $\boldsymbol{\sigma}'$ is expressed as

$$\boldsymbol{\sigma} = \boldsymbol{\sigma}' - \mathbf{m}p \quad , \tag{12}$$

in which $p$ is the pore water pressure, and $\mathbf{m}$ is the second-order identity tensor. Considering the thermal expansion of solid, the effective stress $\boldsymbol{\sigma}'$ can be further refined as

$$\begin{cases} d\boldsymbol{\sigma}' = \mathbb{D}^{ep} : (d\boldsymbol{\varepsilon} - d\boldsymbol{\varepsilon}_T) \\ \boldsymbol{\varepsilon} = \dfrac{1}{2}(\nabla \mathbf{u} + (\nabla \mathbf{u})^T) \quad , \\ \boldsymbol{\varepsilon}_T = \boldsymbol{\alpha}(T - T_0) \end{cases} \tag{13}$$

where $\mathbb{D}^{ep}$, $\boldsymbol{\varepsilon}$, and $\boldsymbol{\varepsilon}_T$ are the fourth-order constitutive tensor, total strain, and thermal strain, respectively. $\boldsymbol{\alpha}$ is the linear thermal expansion coefficient tensor, while $T$ represents the temperature, and $T_0$ denotes the reference temperature.

*2.3 Energy conservation equation*

Heat transfer in solid typically involves three mechanisms: conduction, convection, and radiation. In this paper, last type is not considered. By comparing the magnitudes of different terms in the energy equation, Gelet et al. (2012) found that the hydro-mechanical term's contribution to heat generation was much smaller than the heat transfer term, particularly for geotechnical materials. Therefore, as suggested in Bear and Corapcioglu (1981), McTigue (1986), and Nair et al. (2004), the contribution of hydro-mechanical terms to heat generation is neglected in the energy equation in this study. Assuming uniform temperature between the solid and fluid phases in the mixture and neglecting the convective heat transfer of solid phase, the energy conservation equation can be given as

$$\rho_m C_m \frac{dT}{dt} + \rho_f C_f \mathbf{v}_r \cdot \nabla T + \nabla \cdot (-k_T \nabla T) = Q_T \quad , \tag{14}$$

where $\rho_m C_m = (1-\phi)\rho_s C_s + \phi \rho_f C_f$ is the specific heat capacity of mixture. $\rho_s C_s$ and $\rho_f C_f$ are the heat capacity of solid and pore fluid, respectively. $k_T$ represents the heat conductivity, and $Q_T$ denotes the heat source. The second term in Eq.(14) accounts for the heat contribution of fluid advective transport, and the third term corresponds to the thermal diffusion, in accordance with Fourier's law. It is important to recognize that Eq.(14) excludes the contributions of solid deformation and pressure gradient to the heat flux due to their comparatively minor magnitude (Bear and Corapcioglu, 1981; Gelet et al., 2012; McTigue, 1986).

## 3. ES-FEM with bubble function for THM analysis

### 3.1 Weak forms of governing equations

Considering a quasi-static problem and employing virtual displacement as the trial function, the weak form of the momentum conservation equation is expressed as

$$\mathcal{L}_u = \int_\Omega (\nabla \delta \mathbf{u} : \boldsymbol{\sigma} - \delta \mathbf{u} \cdot \mathbf{b}) d\Omega - \int_{\Gamma_t} \delta \mathbf{u} \cdot \bar{\mathbf{t}} d\Gamma \ , \tag{15}$$

The displacement field in Eq.(15) satisfies the Dirichlet boundary $\Gamma_u$ and Neumann boundary $\Gamma_t$ condition over the material domain $\Omega$ in Eq.(16).

$$\begin{cases} \mathbf{u} = \bar{\mathbf{u}} & \text{on } \Gamma_u \\ \mathbf{n} \cdot \boldsymbol{\sigma} = \bar{\mathbf{t}} & \text{on } \Gamma_t \end{cases} \tag{16}$$

in which $\mathbf{n}$ is the outward vector in $\Gamma_t$.

Utilizing virtual fluid pressure as the trial function and integrating the Eq.(10) over the material domain $\Omega$ via the standard Galerkin method, we can obtain the weak form of mass conservation equation as

$$\mathcal{L}_p = \int_\Omega \left( \delta p \nabla \cdot \mathbf{v} + \delta p \nabla \cdot \mathbf{v}_r - \delta p \cdot 3\alpha \frac{dT}{dt} - \delta p \frac{Q_f}{\rho_f} \right) d\Omega \ . \tag{17}$$

The Darcy velocity can also be written as

$$\mathbf{v}_r = -\frac{\mathbf{k}_f}{\gamma_f} \cdot \nabla p \tag{18}$$

where $\mathbf{k}_f$ and $\gamma_f$ are the permeability matrix and specific weight of pore water, respectively. Substituting Eq.(18) into Eq.(17) and neglecting the pore fluid source term, the weak form is further expressed as

$$\mathcal{L}_p = \int_\Omega \left( \delta p \nabla \cdot \mathbf{v} + \nabla \delta p \cdot \frac{\mathbf{k}_f}{\gamma_f} \cdot \nabla p - \delta p \cdot 3\alpha \frac{dT}{dt} \right) d\Omega - \int_{\Gamma_q} \delta p \cdot \overline{q} \, d\Gamma , \tag{19}$$

where the pore water pressure boundary $\Gamma_p$ and flow flux boundary $\Gamma_q$ satisfy the following Eq.(20).

$$\begin{cases} p = \overline{p} & \text{on } \Gamma_p \\ -\mathbf{n} \cdot \mathbf{v}_r = \overline{q} & \text{on } \Gamma_q \end{cases} \tag{20}$$

Employing temperature field as the trial function and integrating the Eq.(14) via the standard Galerkin method, we can obtain the weak form of energy conservation equation as

$$\mathcal{L}_T = \int_\Omega \left( \rho_m C_m \delta T \cdot \frac{dT}{dt} + \rho_f C_f \delta T \cdot \mathbf{v}_r \cdot \nabla T + \delta T \cdot \nabla \cdot (-k_T \nabla T) - \delta T Q_T \right) d\Omega . \tag{21}$$

Substituting Eq.(18) into Eq.(21) and considering the absence of a heat source, the weak form of energy conservation equation can be rewritten as

$$\mathcal{L}_T = \int_\Omega \left( \rho_m C_m \delta T \cdot \frac{dT}{dt} + \rho_f C_f \delta T \cdot \mathbf{v}_r \cdot \nabla T + k_T \nabla \delta T \cdot \nabla T \right) d\Omega - \int_{\Gamma_h} \delta T \cdot \overline{q}_T \, d\Gamma , \tag{22}$$

in which the temperature boundary $\Gamma_T$ and heat flux boundary $\Gamma_h$ adhere to the expression in Eq.(23).

$$\begin{cases} T = \overline{T} & \text{on } \Gamma_T \\ \mathbf{n} \cdot (k_T \nabla T) = \overline{q}_T & \text{on } \Gamma_h \end{cases} \tag{23}$$

## 3.2 Spatial and temporal discretisation

In conventional FE analysis, the shape function for the displacement field typically requires a higher order than that for the pressure field to satisfy the LBB or inf-sup condition (Babuška and Narasimhan, 1997) and thereby avoid pressure instability. Benefiting from the "softening" effect of bES-FEM, the displacement, pressure, and temperature fields can all be discretized spatially using linear T3 elements, with the interpolated expressions as

$$\begin{cases} \mathbf{u} = \mathbf{N}_u \mathbf{U} \\ p = \mathbf{N}_p \mathbf{P} \\ T = \mathbf{N}_T \mathbf{T} \end{cases} , \tag{24}$$

in which $\mathbf{N}_u$, $\mathbf{N}_p$, and $\mathbf{N}_T$ are the shape function of displacement, pressure, and temperature, respectively. $\mathbf{U}$, $\mathbf{P}$ and $\mathbf{T}$ are nodal displacement vector, nodal fluid pressure vector and nodal temperature vector. Substituting Eq.(24) into Eq.(15), (19), and (22) and eliminating the trial function terms, we can obtain the subsequent spatial discretized equations:

$$\mathbf{KU} + \mathbf{GP} + \mathbf{ST} = \mathbf{F}^u , \quad (25)$$

$$\mathbf{G}^T \frac{d\mathbf{U}}{dt} + \mathbf{HP} + \mathbf{Z}\frac{d\mathbf{T}}{dt} = \mathbf{F}^p , \quad (26)$$

$$\mathbf{J}\frac{d\mathbf{T}}{dt} + \mathbf{WT} = \mathbf{F}^h , \quad (27)$$

where

$$\mathbf{K} = \int_\Omega \left(\mathbf{B}_u^T \mathbf{D} \mathbf{B}_u\right) d\Omega , \quad (28)$$

$$\mathbf{G} = \int_\Omega \left(-\mathbf{B}_u^T \mathbf{m} \mathbf{N}_p\right) d\Omega , \quad (29)$$

$$\mathbf{S} = \int_\Omega \left(-\mathbf{B}_u^T \mathbf{D}\boldsymbol{\alpha} \mathbf{N}_T\right) d\Omega , \quad (30)$$

$$\mathbf{H} = \int_\Omega \left(-\mathbf{B}_p^T \frac{\mathbf{k}_f}{\gamma_f} \mathbf{B}_p\right) d\Omega , \quad (31)$$

$$\mathbf{Z} = \int_\Omega \left(3\alpha \mathbf{N}_p^T \mathbf{N}_T\right) d\Omega , \quad (32)$$

$$\mathbf{J} = \int_\Omega \left(\rho_m C_m \mathbf{N}_T^T \mathbf{N}_T\right) d\Omega , \quad (33)$$

$$\mathbf{W} = \int_\Omega \left(\rho_f C_f \mathbf{N}_T^T \mathbf{v}_r^T \mathbf{B}_T + k_T \mathbf{B}_T^T \mathbf{B}_T\right) d\Omega , \quad (34)$$

$$\mathbf{B}_u = \mathbf{L}_u \mathbf{N}_u = \begin{bmatrix} \mathbf{B}_u^1 & \mathbf{B}_u^2 & \cdots & \mathbf{B}_u^n \end{bmatrix}, \quad \mathbf{L}_u = \begin{bmatrix} \frac{\partial}{\partial x} & 0 & \frac{\partial}{\partial y} \\ 0 & \frac{\partial}{\partial y} & \frac{\partial}{\partial x} \end{bmatrix}^T \quad (35)$$

$$\mathbf{B}_T = \mathbf{L}_T \mathbf{N}_T = \mathbf{B}_p = \mathbf{L}_p \mathbf{N}_p = \begin{bmatrix} \mathbf{B}_p^1 & \mathbf{B}_p^2 & \cdots & \mathbf{B}_p^n \end{bmatrix}, \quad \mathbf{L}_T = \mathbf{L}_p = \begin{bmatrix} \frac{\partial}{\partial x} & \frac{\partial}{\partial y} \end{bmatrix}^T \quad (36)$$

$$\mathbf{F}^u = \int_\Omega \mathbf{N}_u^T \mathbf{b} d\Omega + \int_{\Gamma_t} \mathbf{N}_u^T \overline{\mathbf{t}} d\Gamma , \quad (37)$$

$$\mathbf{F}^p = -\int_{\Gamma_q} \mathbf{N}_p^T \overline{q} d\Gamma , \quad (38)$$

$$\mathbf{F}^h = \int_{\Gamma_h} \mathbf{N}_T^T \overline{q}_T d\Gamma . \quad (39)$$

The study employs a time discretization method based on the trapezoidal rule (Sloan and Abbo, 1999; Smith et al., 2013). Considering that one physical variable changes linearly from the current time step $t_n$ to the next time step $t_{n+1}$, we can integrate the mass and energy conservation equations at the intermediate time step $t_{n+\theta}$, using the variables in Eq.(40), where $0 \leq \theta \leq 1$.

$$\begin{cases} \mathbf{P}_{t_{n+\theta}} = (1-\theta_1)\mathbf{P}_{t_n} + \theta_1 \mathbf{P}_{t_{n+1}} \\ \mathbf{F}^p_{t_{n+\theta}} = (1-\theta_1)\mathbf{F}^p_{t_n} + \theta_1 \mathbf{F}^p_{t_{n+1}} \\ \mathbf{T}_{t_{n+\theta}} = (1-\theta_2)\mathbf{T}_{t_n} + \theta_2 \mathbf{T}_{t_{n+1}} \\ \mathbf{F}^h_{t_{n+\theta}} = (1-\theta_2)\mathbf{F}^h_{t_n} + \theta_2 \mathbf{F}^h_{t_{n+1}} \end{cases} \quad (40)$$

Typically, $\theta$ within the range of 0.5 to 1.0 can ensure stable convergence. In this work, we prefer to set $\theta$ as 1, which corresponds to the backward Euler formulation. Substituting Eq.(40) into Eqs.(26) (27) and integrating over a time increment $\Delta t$, we further derive the following expression:

$$\mathbf{G}^T \mathbf{U}_{t_{n+1}} + \Delta t \mathbf{H} \cdot \mathbf{P}_{t_{n+1}} + \mathbf{Z} \mathbf{T}_{t_{n+1}} = \Delta t \mathbf{F}^p_{t_{n+1}} + \mathbf{G}^T \mathbf{U}_{t_n} + \mathbf{Z} \mathbf{T}_{t_n}, \quad (41)$$

$$\mathbf{J} \mathbf{T}_{t_{n+1}} + \Delta t \mathbf{W} \mathbf{T}_{t_{n+1}} = \Delta t \mathbf{F}^h_{t_{n+1}} + \mathbf{J} \mathbf{T}_{t_n}. \quad (42)$$

Combining Eq.(25), (41), and (42), we can obtain the THM coupled matrix form as

$$\begin{bmatrix} \mathbf{K} & \mathbf{G} & \mathbf{S} \\ \mathbf{G}^T & \Delta t \mathbf{H} & \mathbf{Z} \\ \mathbf{0} & \mathbf{0} & \mathbf{J} + \Delta t \mathbf{W} \end{bmatrix} \begin{bmatrix} \mathbf{U}_{t_{n+1}} \\ \mathbf{P}_{t_{n+1}} \\ \mathbf{T}_{t_{n+1}} \end{bmatrix} = \begin{bmatrix} \mathbf{F}^u_{t_{n+1}} \\ \Delta t \mathbf{F}^p_{t_{n+1}} + \mathbf{G}^T \mathbf{U}_{t_n} + \mathbf{Z} \mathbf{T}_{t_n} \\ \Delta t \mathbf{F}^h_{t_{n+1}} + \mathbf{J} \mathbf{T}_{t_n} \end{bmatrix}. \quad (43)$$

### 3.3 Edge-based smoothed technique with bubble function

Compared to conventional FEM, ES-FEM provides solutions with higher accuracy and demonstrates insensitivity to mesh distortion due to the "softening effects" of the smoothing domain (Liu and Trung, 2016). Despite the considerable promise of ES-FEM, it is not fully volumetric locking-free for nearly incompressible materials (Nguyen-Xuan and Liu, 2013). Fortunately, it can be conveniently improved to the bES-FEM formulation to fulfil the inf-sup condition, making it suitable for incompressible materials. As illustrated in Fig. 1, the smoothing domain is established by connecting two endpoints of an edge with the centroids of the adjacent elements. This method enriches each T3 element by supplementing a new central node (known as the bubble point), thereby introducing two new degrees of freedom. This kind of elements, known as MINI elements, were first proposed in Arnold et al. (1984) for Stokes flow problems. Such elements are noted to satisfy the inf-sup condition (Brezzi and Fortin, 2012), offering a notably simple and effective means for solving incompressible limit. Notably, Fig. 1 only illustrates the method with triangular elements as examples, yet it is also applicable to quadrilateral and other polygonal elements. Accordingly, a cubic bubble shape function is supplemented into the shape functions of T3 elements in the current bES-FEM formulation, as shown in Eq.(44).

$$\mathbf{u} = \sum_{i=1}^{3} \begin{bmatrix} N_u^i(\mathbf{x}) & 0 \\ 0 & N_u^i(\mathbf{x}) \end{bmatrix} \mathbf{U}_i + \begin{bmatrix} N_u^b(\mathbf{x}) & 0 \\ 0 & N_u^b(\mathbf{x}) \end{bmatrix} \mathbf{U}_b \ , \tag{44}$$

where $N_u^i(\mathbf{x})$ and $N_u^b(\mathbf{x})$ are the shape function of T3 elements and supplemented bubble shape function at the centroid of T3 elements, respectively. $\mathbf{U}_b$ is the displacement of the complementary bubble point, and $N_u^b$ can be expressed as

$$N_u^b = 27 L_1 L_2 L_3 \quad (L_i = a_i + b_i x + c_i y; \ i = 1, 2, 3) \ , \tag{45}$$

in which $L_i$ represents the area coordinates with $a_i, b_i$ and $c_i$ satisfying the Eq.(46).

$$a_i = \frac{1}{2A_e}(x_j y_k - x_k y_j), \quad b_i = \frac{1}{2A_e}(y_j - y_k), \quad c_i = \frac{1}{2A_e}(x_k - x_j) \ , \tag{46}$$

where $A_e$ is the area of the triangular element with $i, j, k$ denoting the cyclic permutation of its three nodes. It is noted that the bubble function within the triangular element domain $\Omega_e$ adheres to the following properties:

$$N_u^b > 0, \forall \mathbf{x} \in \Omega_e; \quad N_u^b = 0, \forall \mathbf{x} \in \partial \Omega_e; \quad N_u^b(\mathbf{x}_{\text{centroid}}) = 1.0 \ . \tag{47}$$

The constant smoothing strain $\boldsymbol{\varepsilon}$ within the smoothing domain $\Omega_k$ can be given as

$$\boldsymbol{\varepsilon} = \frac{1}{A_e} \int_{\Omega_k} \mathbf{B}_u \mathbf{u} d\Omega = \sum_{I \in \Psi^k} \mathbf{B}_I \mathbf{U}_I \ , \tag{48}$$

where $I$ is one node in the nodal set $\Psi^k$ supporting to the smoothing domain $\Omega_k$. By substituting Eq.(35) and (44) into Eq.(48), we can get the expression of $\mathbf{B}_I$ in the following matrix form:

$$\mathbf{B}_I = \begin{bmatrix} b_{Ix}(\mathbf{x}_k) & 0 \\ 0 & b_{Iy}(\mathbf{x}_k) \\ b_{Iy}(\mathbf{x}_k) & b_{Ix}(\mathbf{x}_k) \end{bmatrix} \ , \tag{49}$$

with

$$b_{Ih}(\mathbf{x}_k) = \frac{1}{A_e}\int_{\Gamma_k} N_I n_h \mathrm{d}\Gamma = \frac{1}{A_e}\sum_{b=1}^{n_{eg}} N_I n_h l_b, \quad (h = x, y). \tag{50}$$

In Eq.(50), $n_{eg}$ is the number of boundary segments with $n_h$ representing the components of the unit outward vector on these segments. $N_I$ denotes the value of the shape function at gauss points, while $l_b$ represents the length of boundary segments. In this work, the T3 element with a supplemental bubble point using smooth integration is also termed as bES-T3. In essence, the core concept of strain smoothing in bES-T3 involves replacing $\mathbf{B}_u$ with $\mathbf{B}_I$ accounting for the bubble point. Similarly, the smoothed gradient matrices $\mathbf{B}_p$ for the pressure field and $\mathbf{B}_T$ for the temperature field can also be easily processed using the method.

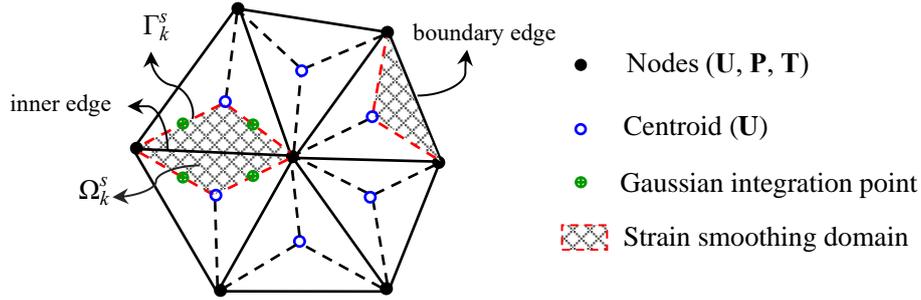

Fig. 1 Diagram of triangular mesh and smoothing domains in bES-FEM-T3

*3.4 SUPG scheme*

Within the conventional FEM framework, the application of the standard Galerkin method ensures the symmetry of the stiffness matrix and generally provides a satisfactory approximation (Brooks and Hughes, 1982). However, spatial oscillations may arise from the inherent asymmetry of the stiffness matrix in the energy conservation equation, especially in convection-dominated heat transfer scenarios with high-velocity fluids (Brooks and Hughes, 1982; Cacace and Jacquey, 2017). These oscillations are attributable to the standard Galerkin FEM framework treating upstream and downstream nodes with equal weight, despite the physical reality necessitating a greater influence for the upstream node. To mitigate these instabilities, this work adopts the Streamline Upwind Petrov-Galerkin (SUPG) scheme in the energy conservation equation, which modifies the test function along the streamline direction, thereby increasing the weight of the upstream nodes. Contrasting the standard Galerkin method with identical trial and shape functions, the Petrov-Galerkin method supplements an artificial stabilization term related to the velocity field into the trial function, which is expressed as ($\mathbf{v}_r$ is assumed to have the format of a column vector)

$$\bar{\mathbf{N}}_T = \mathbf{N}_T + \tau_s \mathbf{v}_r^T (\nabla \mathbf{N}_T) = \mathbf{N}_T + \tau_s \mathbf{v}_r^T \mathbf{B}_T. \tag{51}$$

In the current SUPG scheme, the weak form in Eq.(22) can thus be reformulated as

$$\mathcal{L}_T = \int_\Omega \left( \rho_m C_m \bar{\mathbf{N}}_T^T \frac{dT}{dt} + \rho_f C_f \bar{\mathbf{N}}_T^T \cdot \mathbf{v}_r \cdot \nabla T \right) d\Omega + \int_\Omega \left( k_T \nabla \bar{\mathbf{N}}_T \cdot \nabla T \right) d\Omega - \int_{\Gamma_h} \bar{\mathbf{N}}_T^T \bar{q}_T d\Gamma = 0, \tag{52}$$

where for T3 element, $\nabla \bar{\mathbf{N}}_T$ is also equal to $\mathbf{B}_T$, as $\mathbf{v}_r^T \mathbf{B}_T$ is indeed a constant over one element. Accordingly, the matrix form of the energy equation after spatial discretization can be written as

$$(\mathbf{J} + \mathbf{J}_s) \frac{d\mathbf{T}}{dt} + (\mathbf{W} + \mathbf{W}_s) \mathbf{T} = \mathbf{F}^h + \mathbf{F}^{hs}, \tag{53}$$

in which

$$\mathbf{J}_s = \int_\Omega \left( \rho_m C_m \tau_s \mathbf{B}_T^T \mathbf{v}_r \mathbf{N}_T \right) d\Omega, \tag{54}$$

$$\mathbf{W}_s = \int_\Omega \left( \rho_f C_f \tau_s \mathbf{B}_T^T \mathbf{v}_r \mathbf{v}_r^T \mathbf{B}_T \right) d\Omega, \tag{55}$$

$$\mathbf{F}^{hs} = \int_{\Gamma_h} \tau_s \mathbf{B}_T^T \mathbf{v}_r \bar{q}_T d\Gamma. \tag{56}$$

The selection of the stabilization parameter $\tau_s$ is problem-dependent, markedly affecting the stability and precision of solutions. In our work, we utilize an expression of $\tau_s$ recommended in Brooks and Hughes (1982) and Cui and Wong (2022), given as

$$\tau_s = \frac{\bar{\xi} v_\xi h_\xi + \bar{\eta} v_\eta h_\eta}{2 |\mathbf{v}_r|^2}, \tag{57}$$

in which

$$\begin{aligned} \bar{\xi} &= \coth(Pe_\xi) - 1/Pe_\xi, & \bar{\eta} &= \coth(Pe_\eta) - 1/Pe_\eta \\ Pe_\xi &= v_\xi h_\xi / 2\kappa_T, & Pe_\eta &= v_\eta h_\eta / 2\kappa_T \end{aligned} \tag{58}$$

The parameters $v_\xi$ and $v_\eta$ in Eq.(58) denote the nodal velocity components along the local coordinate axis of the element, while $h_\xi$ and $h_\eta$ represent the corresponding characteristic lengths, as shown in Fig. 3.4. of Brooks and Hughes (1982). $\kappa_T$ is known as the heat diffusivity coefficient, which can be calculated as $\kappa_T = k_T / (\rho_m C_m)$. Adopting the backward Euler method for time discretization as described in Section 3.2, the formulation of the coupling THM matrix incorporating the SUPG scheme is subsequently derived as

$$\begin{bmatrix} \mathbf{K} & \mathbf{G} & \mathbf{S} \\ \mathbf{G}^T & \Delta t \mathbf{H} & \mathbf{Z} \\ \mathbf{0} & \mathbf{0} & (\mathbf{J}+\mathbf{J}_s)+\Delta t(\mathbf{W}+\mathbf{W}_s) \end{bmatrix} \begin{bmatrix} \mathbf{U}_{t_{n+1}} \\ \mathbf{P}_{t_{n+1}} \\ \mathbf{T}_{t_{n+1}} \end{bmatrix} = \begin{bmatrix} \mathbf{F}^u_{t_{n+1}} \\ \Delta t \mathbf{F}^p_{t_{n+1}} + \mathbf{G}^T \mathbf{U}_{t_n} + \mathbf{Z}\mathbf{T}_{t_n} \\ \Delta t \left(\mathbf{F}^h_{t_{n+1}} + \mathbf{F}^{hs}_{t_{n+1}}\right) + (\mathbf{J}+\mathbf{J}_s)\mathbf{T}_{t_n} \end{bmatrix}. \quad (59)$$

## 4. Verification of the proposed approach

To validate the accuracy of the bES-FEM formulation with SUPG stabilization for THM coupled problems, a series of benchmark tests ranging from simple to complex were carried out. First, a one-dimensional (1D) Terzaghi's consolidation model was conducted to demonstrate the method's superiority in simulating nearly undrained conditions. Then, a footing case was performed to highlight the current bES-FEM formulation's robustness to mesh distortion and its advantage in overcoming pore pressure instability. Furthermore, a classical steady-state convective heat transfer model was simulated to demonstrate the effectiveness of the SUPG scheme in eliminating spurious oscillations arising from convection-dominated problems. Finally, two thermo-elastic consolidation cases involving saturated soil were modelled to demonstrate the capacity of the present formulation in addressing coupled THM problem.

### 4.1 1D Terzaghi's consolidation

We considered a Terzaghi's 1D consolidation model for a saturated elastic soil, with a height of 1 m and a width of 0.06 m, as illustrated in Fig. 2(a). A uniform mesh with 711 nodes and 1208 elements was used. The top boundary of the model was set as a free drainage boundary, while the remaining three boundaries were designated as impervious boundaries. To demonstrate the solution precision and convergence of the bES, we first compared the numerical results obtained from bES with those from T3, NS, and ES. The model had an elastic modulus of $E = 10^4$ kPa and a Poisson's ratio of 0.3. To simulate the nearly undrained limit condition, the permeability was set as $k_f = 10^{-9}$ m/s, while the unit weight of fluid is 10 kN/m³. A uniform load $P_0$ = 10 kPa was instantaneously applied on the top boundary at the first step and kept constant for a total consolidation period of 1.5e6 s with a consistent time step of 150 s. For convenient comparison with the analytical solution (Terzaghi, 1943), a dimensionless time is defined in Eq.(60).

$$T_v = \frac{k_f E (1-\nu) t}{\gamma_f H^2 (1+\nu)(1-2\nu)} \tag{60}$$

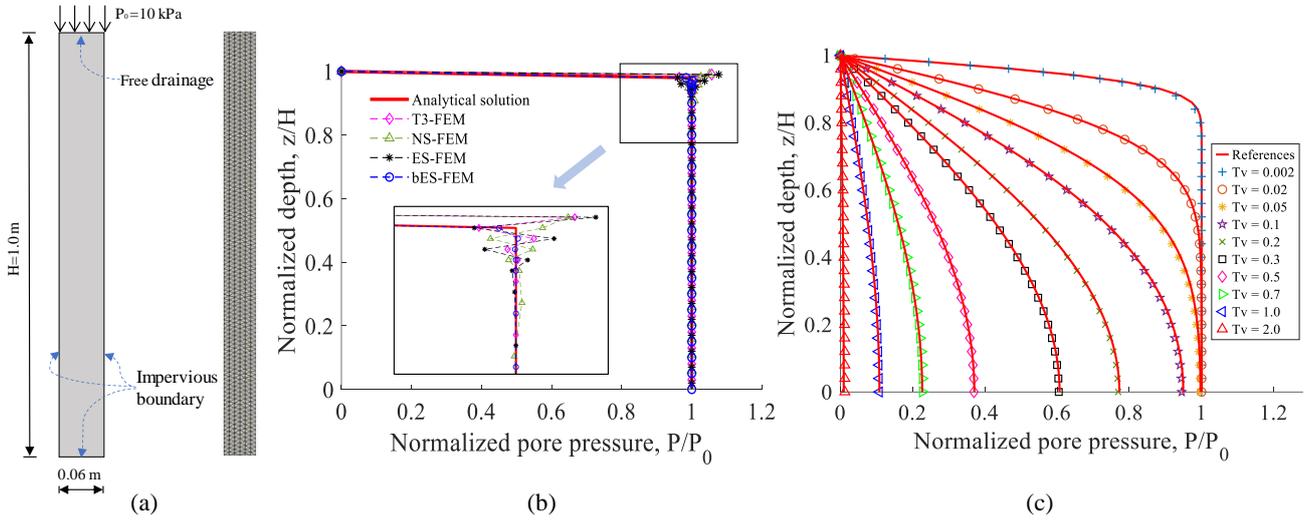

Fig. 2 (a) 1D Terzaghi's model with mesh; (b) distribution of normalized pore pressure with respect to normalized depth with different numerical methods; (c) distribution of normalized pore pressure with respect to normalized depth at various consolidation times

Fig. 2 (b) depicted the distribution of normalized pore pressure versus normalized depth using various numerical methods at the initial consolidation time. It was observed that the numerical results obtained from T3, NS, and ES exhibited significant oscillations at the loading end. In contrast, the bES can produce more stable and accurate pore pressure solutions, highlighting its superiority in overcoming pore pressure oscillations caused by nearly undrained or nearly incompressible conditions. Fig. 2(c) compared the numerical results with the analytical solutions at different consolidation stages. The close agreement between the two demonstrated the excellent performance and capability of the current bES-FEM formulation in handling problems with low permeability coefficients.

*4.2 Footing consolidation of saturated Mohr–Coulomb soil*

To validate the enhanced performance of the current bES-FEM formulation in addressing pore pressure instability and mesh distortion, a flexible strip footing consolidation case featuring hydro-mechanical (HM) coupling was examined. For comparative analysis, two mesh configurations, a regular T3 mesh and a distorted mesh with bES-T3, were used, as shown in Fig. 3. The regular mesh consisted of 483 nodes and 888 elements, while the distorted mesh comprised 995 nodes and 1894 elements. Both mesh configurations had identical boundary conditions and material parameters, with the footing's half-width prescribed at 3 meters ($a = 3$m). The boundary conditions were presented in Fig. 3, while material parameters were provided in Table 1. A uniform load of $P_0 = 100$ kPa was applied to the footing over a period characterized by

$$T_v = \frac{Ekt}{2\gamma_w(1+v)(1-2v)a^2}, \qquad (61)$$

where $t$ represents the real physical time, and $T_v$ denotes the dimensionless time. The load linearly increased from 0.0 to $P_0$ over a period of $T_v = 0.01$ using 10 steps and then remained constant. The time steps then gradually increased by a factor of 1.1 until reaching $T_v = 2.0$.

Table 1 Mohr-Coulomb material parameters

| Parameter | Value | Parameter | Value |
|---|---|---|---|
| Elastic modulus | $E = 2000$ kPa | Cohesion | $c = 10$ kPa |
| Poisson's ratio | $v = 0.3$ | Permeability | $k = 10^{-5}$ m/day |
| Friction angle | $\theta = 20°$ | Unit weight of water | $\gamma_f = 10$ kN/m$^3$ |
| Dilation angle | $\varphi = 20°$ | | |

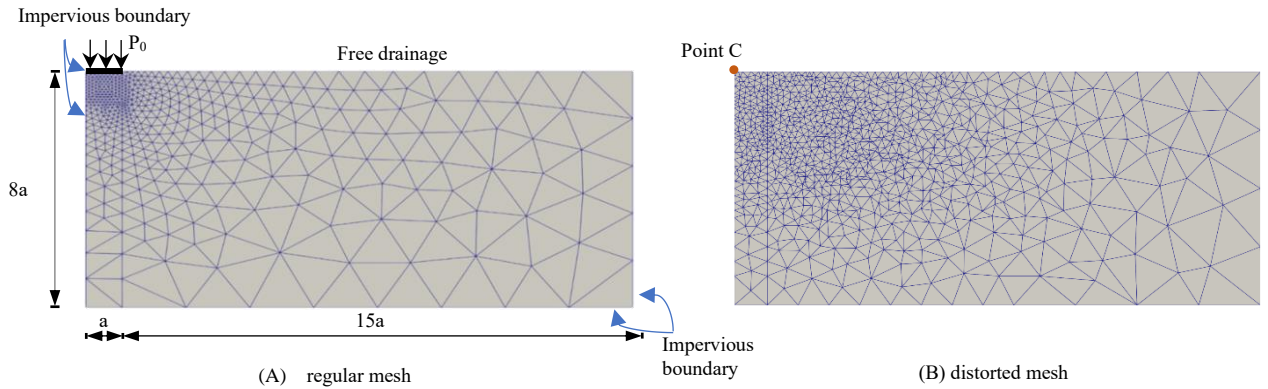

Fig. 3 Strip footing consolidation case with (A) regular T3 mesh and (B) distorted mesh with bES-T3

Fig. 4 displayed the evolution of vertical displacement and pore pressure at point C, clearly showing that bES-T3 produces results that closely matching the reference solution. The consistency of results from distorted meshes with those from regular meshes further demonstrated the low mesh dependency of bES-T3. Using regular meshes, Fig. 5 compared the contours of horizontal displacement, shear stress, and pore pressure fields obtained using bES-FEM with those derived from other methods, namely T3-FEM, NS-FEM, ES-FEM, and T6-FEM. It was clearly observed that T3-FEM, NS-FEM, and ES-FEM all yielded pore pressure field distributions with spurious oscillations, and NS-FEM also produced significant sawtooth horizontal displacement. Such oscillations can be attributed to the failure to satisfy the inf-sup condition during small timesteps or when the material is nearing the undrained limit (Sun, 2015). In contrast, bES-FEM can obtain stable and accurate displacement and pore pressure field distributions. It has been demonstrated that the integration of bubble functions provides an effective numerical stabilization remedy for the inf-sup condition deficiency in bES-FEM (Lamichhane, 2009; Nguyen-Xuan and Liu, 2013), allowing for the simulation of nearly incompressible materials. Notably, bES-FEM not only achieves numerical stability comparable to T6-FEM but also offers an advantage in terms of reduced computational time relative to higher-order elements like T6, as detailed in Table 2. Consequently, bES-FEM can be considered as a promising numerical method that combines stability in pore pressure simulation with relatively higher computational efficiency.

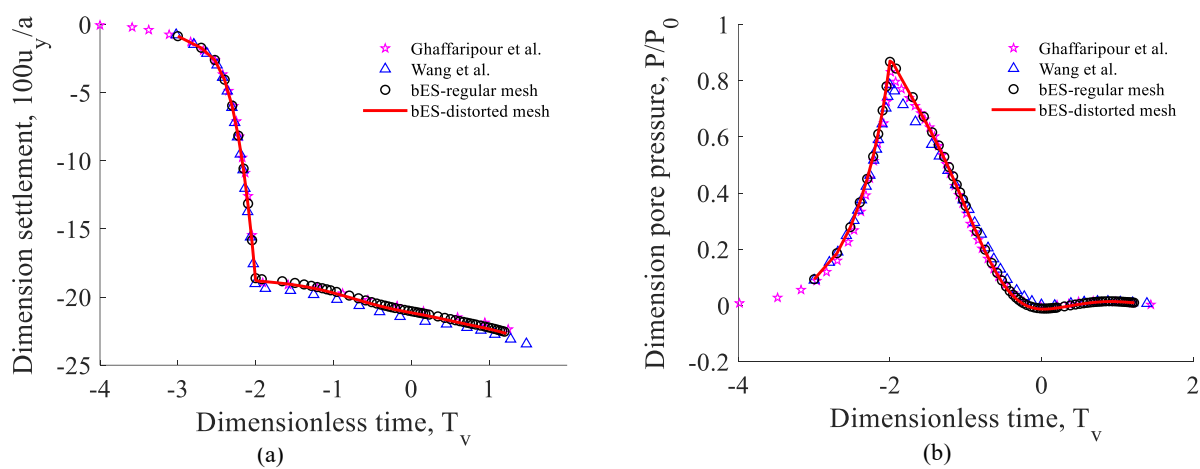

Fig. 4 (a) Variation of dimensionless vertical displacement and (b) dimensionless pressure at point C

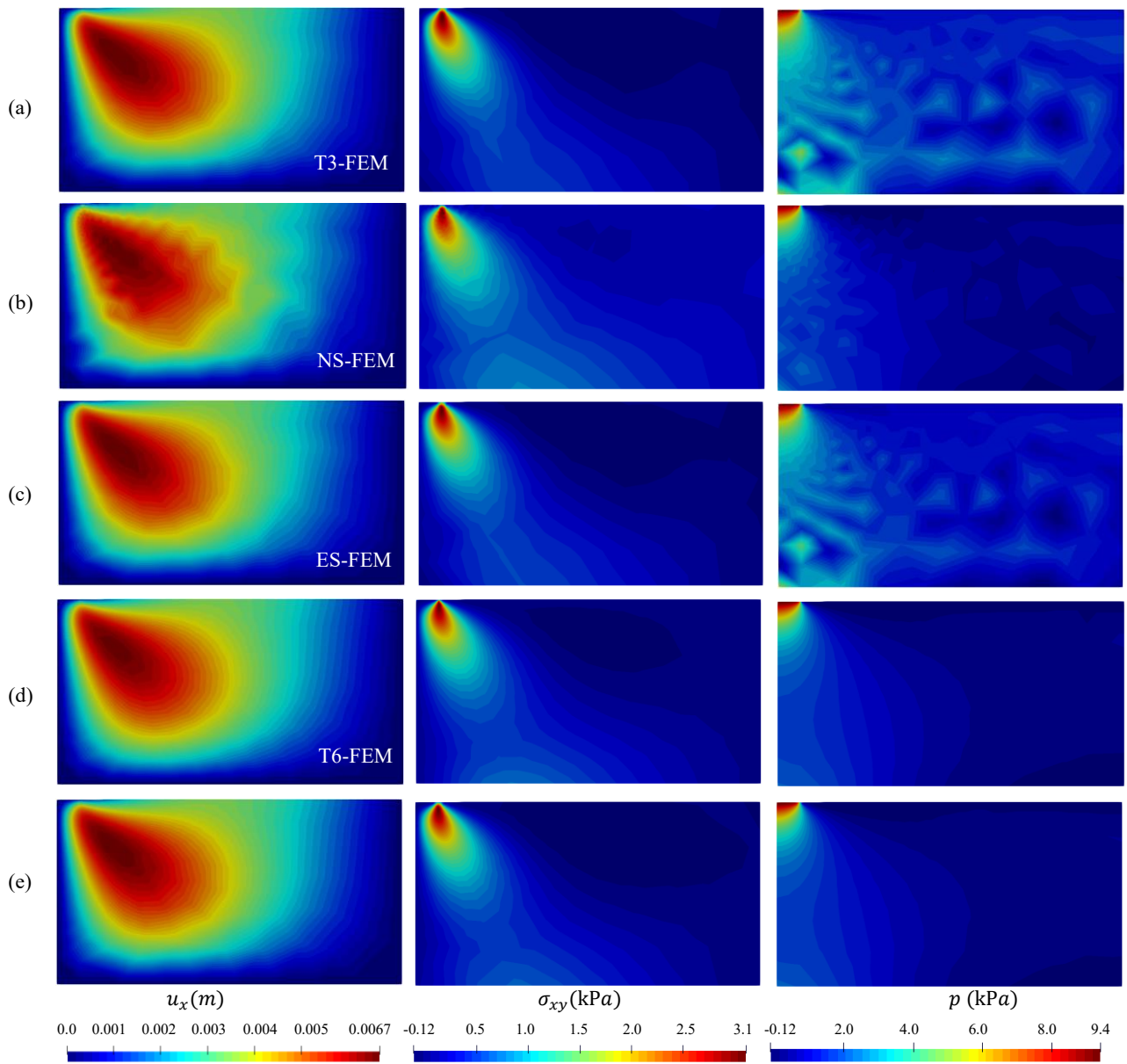

Fig. 5 Contour of horizontal displacement, shear stress and pore pressure at the first step with (a) T3-FEM; (b) NS-FEM; (c) ES-FEM; (d) T6-FEM; (e) bES-FEM

Table 2 Comparison of computational time and efficiency between different methods at $T_v=10$

|  | T3-FEM | NS-FEM | ES-FEM | bES-FEM | T6-FEM |
|---|---|---|---|---|---|
| Computational time (s) | 1.309e5 | 1.903e5 | 1.846e5 | 8.00e5 | 1.11e6 |
| Average number of iterations per step | 5.65 | 6.01 | 6.56 | 7.09 | 7.11 |

*4.3 Steady-state convective heat transfer*

Spatial oscillations are reported to be likely to occur in convection-dominated scenarios with high Péclet and Reynolds numbers (Brooks and Hughes, 1982; Cui and Wong, 2022). To illustrate the efficacy of the integrated SUPG scheme in mitigating these oscillations, a two-dimensional steady-state heat transfer case was examined in this subsection. Fig. 6 presents a porous medium domain with a length of 2.0 meters long and a width of 1.5 meters. A uniform mesh with 3685 nodes and 7068 elements was adopted, and a total period of 4000 s with a consistent time step of 5 s was employed. A fluid pore pressure of 20 kPa was applied at the left boundary, and zero pore pressure was applied at the right boundary to establish a steady-state flow. The fluid was assumed to be incompressible with a viscosity of 0.001 Pa·s, and the other material parameters were listed in Table 3. For a steady-state flow with a permeability coefficient of $10^{-11} m^2$, the fluid velocity is 0.0001 m/s. Enhanced fluid velocities can be realized by elevating the permeability coefficient or intensifying the hydraulic gradient. The model was initially set at a temperature of 0℃, with a nonzero temperature condition applied along the Y-axis at the left boundary, satisfying the following expression

$$T_{left}(y) = \begin{cases} 0 & \text{for } |y| > 0.25m \\ 10 \cdot (2.5 - 10y) & \text{for } 0.15m \leq |y| \leq 0.25m \\ 10 & \text{for } |y| < 0.15m \end{cases}. \quad (62)$$

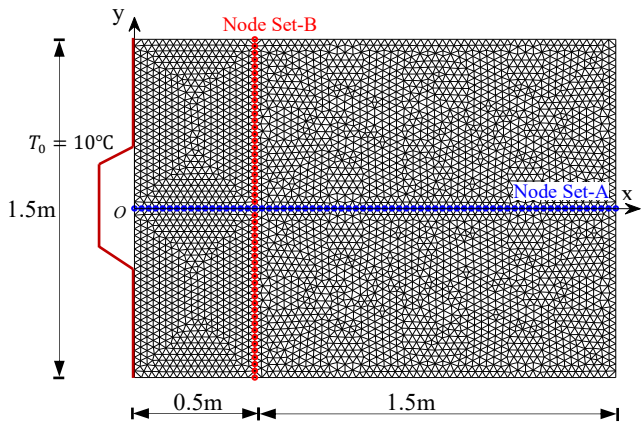

Table 3 Material parameters in the heat transfer case

| Parameter | Value |
| --- | --- |
| Porosity | $\phi = 0.1$ |
| Permeability | $k_f = 10^{-11}$ m² |
| Fluid density | $\rho_f = 1000$ kg/m³ |
| Solid density | $\rho_s = 2000$ kg/m³ |
| Fluid heat capacity | $c_f = 1100$ J/(kg · K) |
| Solid heat capacity | $c_s = 250$ J/(kg · K) |
| Solid thermal conductivity | $k_T^s = 2.0$ W/(m · K) |
| Fluid thermal conductivity | $k_T^f = 0.5$ W/(m · K) |

Fig. 6 Geometric model with mesh and temperature boundary

To verify the capability of bES-FEM formulation in simulating heat transfer, the temperature distribution was tracked at various times in Node Set-A (horizontal) and Set-B (vertical), as marked in Fig. 6, under a flow velocity of 0.0001 m/s. Fig. 7 showed that the simulated temperature distribution closely corresponded with that from the analytical solution provided in Kolditz et al. (2015). In addition to these findings, the study also investigated temperature field distributions at three different flow velocities: 0.0001 m/s, 0.01 m/s, and 1.0 m/s. Fig. 8 revealed that the numerical temperature fields were in close agreement with the analytical solutions at the low flow velocity (0.0001m/s), regardless of the application of SUPG stabilization. However, in the absence of SUPG stabilization, higher flow velocities led to significant numerical oscillations in the temperature field. These high velocities, which correspond to elevated Péclet numbers, resulted in distinct undulations within the heat wavefront region and a notably rough temperature distribution. Fortunately, the implementation of SUPG stabilization in this study effectively alleviated these numerical oscillations, leading to a smoother temperature field.

To quantitatively assess the differences between the implementations with and without SUPG technique, a concept of the mean error (denoted by $c_T$) between numerical and analytical temperatures is introduced, as defined in Eq.(63). In this expression, $T_i^s$ and $T_i^a$ represent the temperatures obtained at the $i$th node of a set comprising $n$ nodes, calculated via simulation and analytical methods, respectively. Fig. 9 presented the mean temperature errors at a specific time under various flow velocities. At low flow velocities, similarly small errors were achieved regardless of the application of SUPG. As flow velocity increased, the mean error between the simulated and analytical temperatures rose. It was observed that SUPG stabilization yielded reduced temperature errors at high flow velocities compared to the non-SUPG method. Additionally, when the flow velocity surpassed 0.1 m/s, the temperature error with SUPG scheme did not experience an abrupt surge as the flow velocity increased.

$$c_T = \frac{1}{n}\sum_{i=1}^{n}\left|T_i^s - T_i^a\right| \tag{63}$$

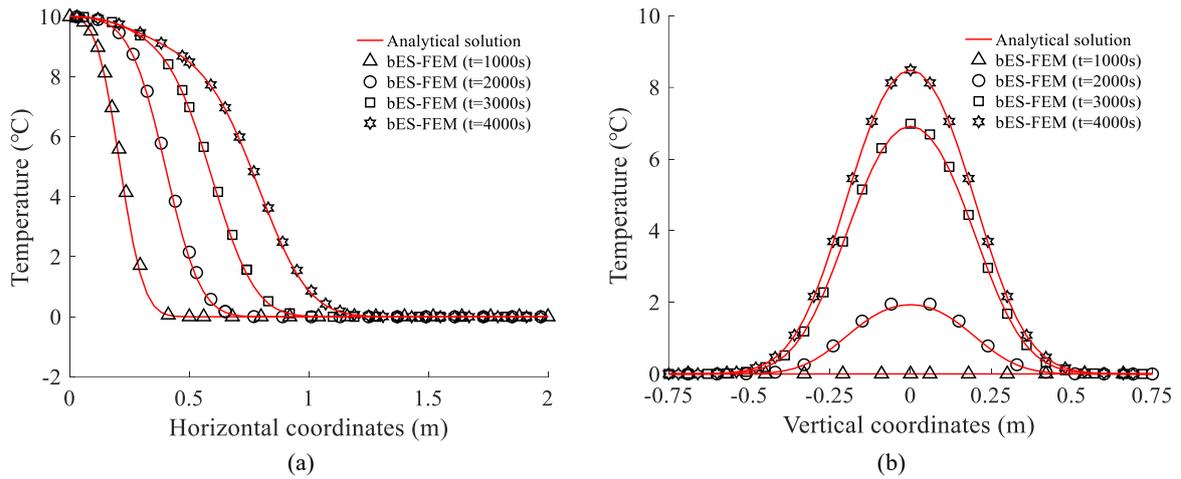

Fig. 7 Comparison of temperatures from analytical and numerical solutions at (a) Node Set-A and (b) Node Set-B

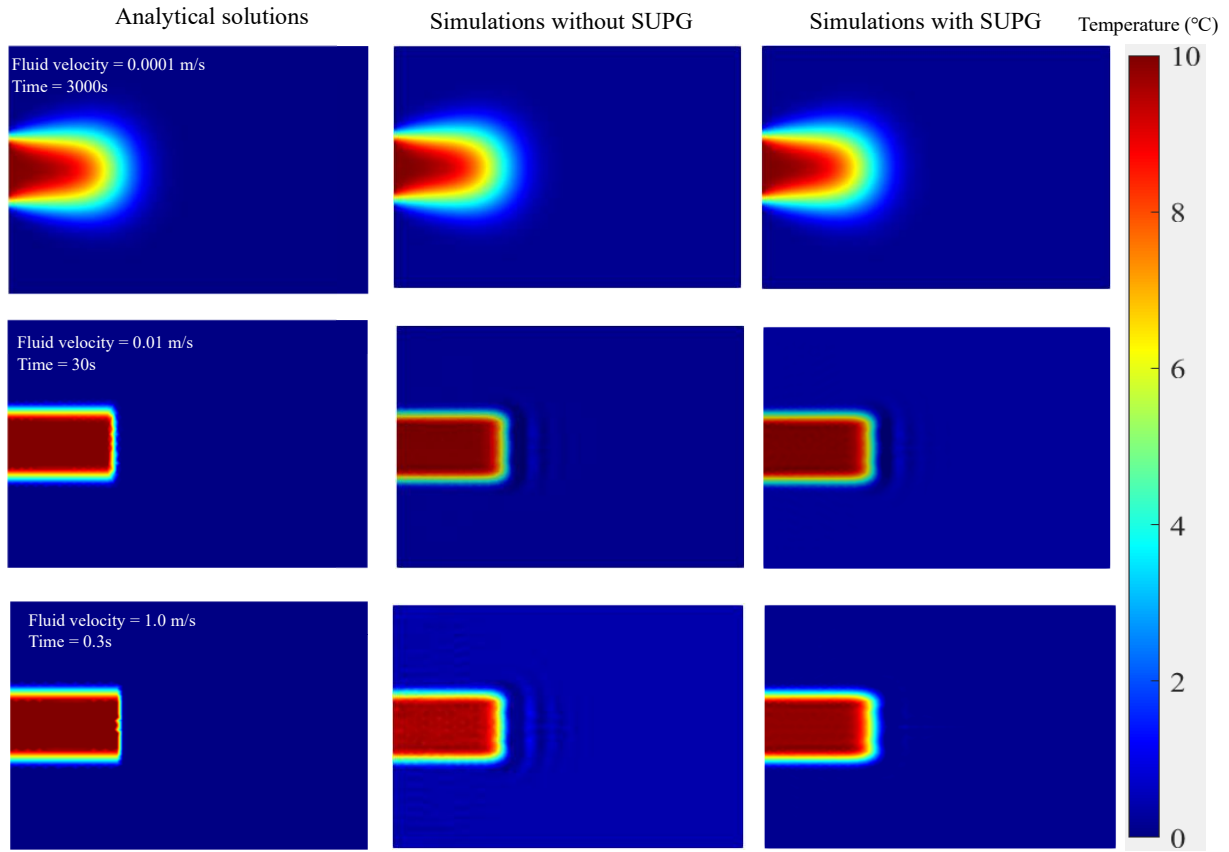

Fig. 8 Comparison of analytical and numerical temperature contour for bES-FEM with/without SUPG

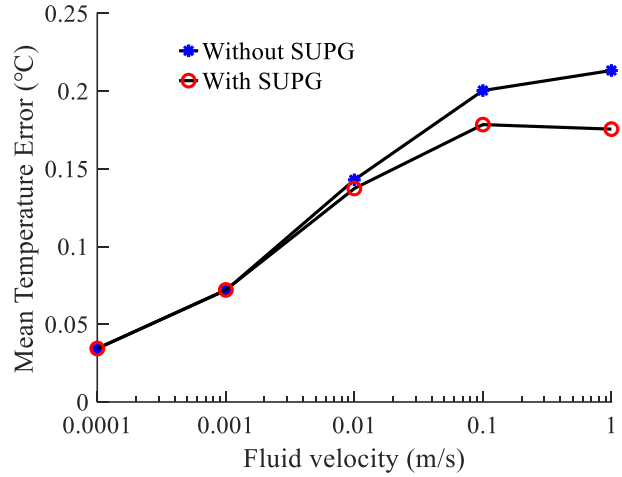

Fig. 9 Comparison of errors for bES-FEM with and without SUPG

*4.4 Thermo-elastic consolidation of half-space*

To demonstrate the superior performance of bES-FEM formulation for coupled THM problems, a half-space thermo-elastic consolidation model from Cheng (2016) was conducted and compared with analytical solutions (McTigue, 1986). Considering the semi-infinite nature of this case, a sufficiently extended plane strain model, depicted by the shaded region in Fig. 10 was created for analysis. A uniform mesh comprising 1205 nodes and 1820 elements was utilized, and the simulation was conducted for a period of 43200 s with a constant time step of 432 s. The left and right boundaries were treated as symmetrical displacement boundaries with impermeable conditions, while the top boundary was designed as a free-draining surface with a constant temperature($T$ = 50℃) to expedite the consolidation process. For simplicity, both the initial thermal field and pore pressure distribution were assumed to be zero, without considering any heat flux and fluid flux. During consolidation, the variations in temperature, pore pressure, and vertical displacement were systematically monitored at five equidistant points labeled A to E in Fig. 10 , each separated by 10 meters. The material parameters were provided in Table 3.

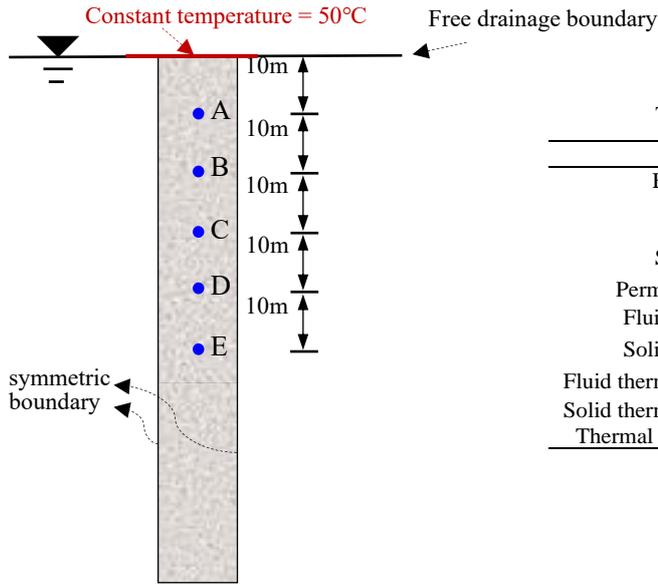

Fig. 10 Half-space model with boundary conditions

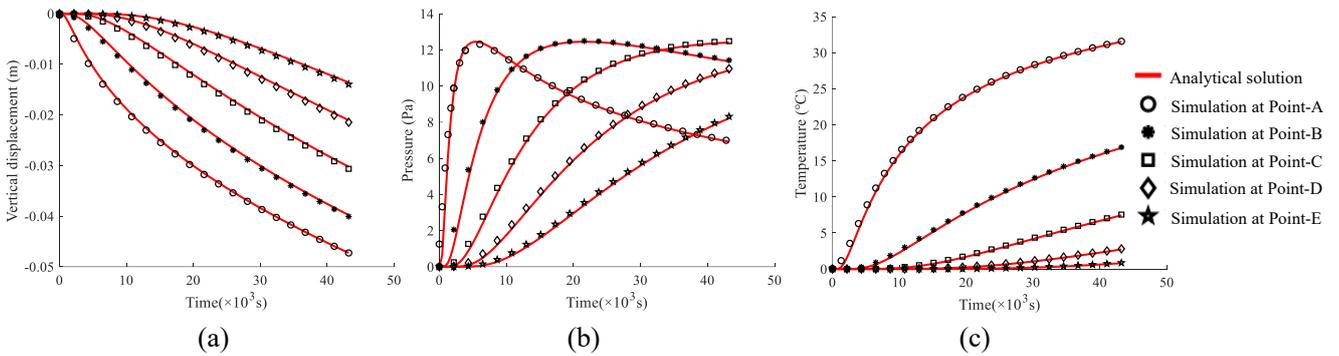

Fig. 11 Comparison of numerical and analytic solutions in (a)vertical displacement, (b)pore pressure, and (c) temperature

Fig. 11 illustrated the temporal evolution of vertical displacement, pore pressure, and temperature at the five measurement points during thermal consolidation. The results demonstrated that the numerical solution agreed well with the analytical solutions, confirming the high accuracy of the current bES-FEM formulation in addressing THM coupling problems. Additionally, it was observed that during the early stages of consolidation, the pore pressure at points A and B gradually increased to a peak before dissipating. At regions close to the free drainage boundary, the rise and subsequent dissipation of pore pressure are more rapid, with an accompanying accelerated temperature increase, aligning with empirical observations. To highlight the robustness of the current bES-FEM formulation in addressing the undrained limit, four supplementary models were created using varying permeability coefficients, specifically 3.89e-6, 9.72e-7, 2.43e7 and 3.89e-8, while keeping all other parameters unchanged. Fig. 12 depicted the temporal variations in excess pore pressure and vertical displacement at point A under different permeability coefficients. It was observed that an increase in permeability coefficient accelerated the rise of excess pore water pressure to its peak, while the peak value itself was reduced. Furthermore, a higher permeability coefficient leaded to a faster dissipation of pore pressure, resulting in smaller vertical displacements and pore pressures. The strong agreement between the numerical and analytical results further corroborated the efficacy and precision of the present bES-FEM formulation for tackling issues of low permeability.

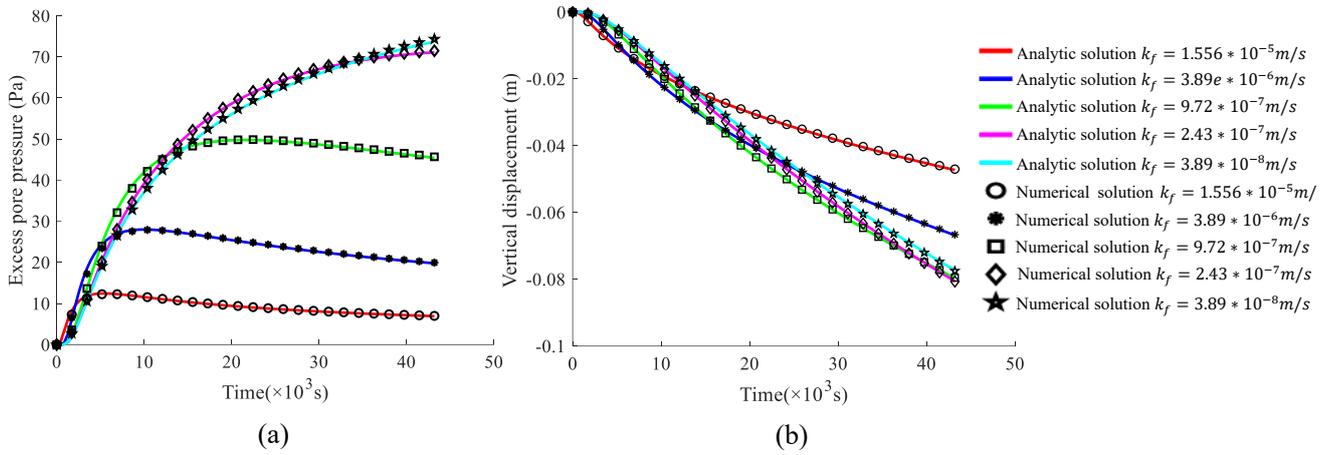

Fig. 12 Comparison of numerical and analytic solutions in (a)excess pore pressure and (b)vertical displacement at point A

*4.5 Thermo-elastic loading consolidation of soil columns*

The one-dimensional thermo-elastic consolidation (Aboustit et al., 1982) has been modelled as a classic engineering paradigm to extensively understand the simultaneous effects of in-situ stress and thermal gradients on strata during energy extraction processes, including underground coal liquefaction and in-situ oil shale retorting. Accordingly, this study also utilized the bES-FEM for a comprehensive analysis of such an example. This numerical example was similar to that outlined in Section 4.4, except that a vertical pressure of 1.0 Pa was applied to the top surface of the soil column (see Fig. 13(a)). A uniform mesh consisting of 396 nodes and 700 elements was employed. To ensure computational accuracy while improving overall computational efficiency, a consistent time step of 1s was used for the first 1000 steps, followed by a time step of 10 s for the remaining 7000 steps. The initial thermal and pore pressure conditions were initialized to zero, and free drainage was permitted exclusively at the upper boundary. To the authors' knowledge, no analytical solutions of this model are available. Nonetheless, researchers have persistently investigated the THM coupling effects in the one-dimensional non-isothermal consolidation problem (Aboustit et al., 1985; Cui et al., 2018; Lewis et al., 1986; Noorishad et al., 1984). The material parameters suggested by Cui et al. (2018), as detailed in Table 5, were utilized,. The temporal variation of temperature and excess pore pressure at Point A, situated one meter below the surface, was systematically recorded.

Table 5 Material parameters for thermo-elastic consolidation

| Properties | Symbol | Unit | Value |
| --- | --- | --- | --- |
| Elastic modulus | $E$ | Pa | 6000 |
| Poisson's ratio | $v$ | - | 0.4 |
| Porosity | $n$ | - | 0.25 |
| Specific weight | $\gamma_f$ | kN/m³ | 9.8 |
| Permeability coefficient | $k_f$ | m/s | 3.92e-2 |
| Fluid thermal capacity | $\rho_f c_f$ | kJ/(m³ K) | 167.2 |
| Solid thermal capacity | $\rho_s c_s$ | kJ/(m³ K) | 167.2 |
| Fluid thermal expansion coefficient | $\alpha_f$ | 1/K | 3.0e-7 |
| Solid thermal expansion coefficient | $\alpha_s$ | 1/K | 3.0e-7 |
| Thermal conductivity coefficient | $k_T$ | kW/(m·K) | 0.836 |

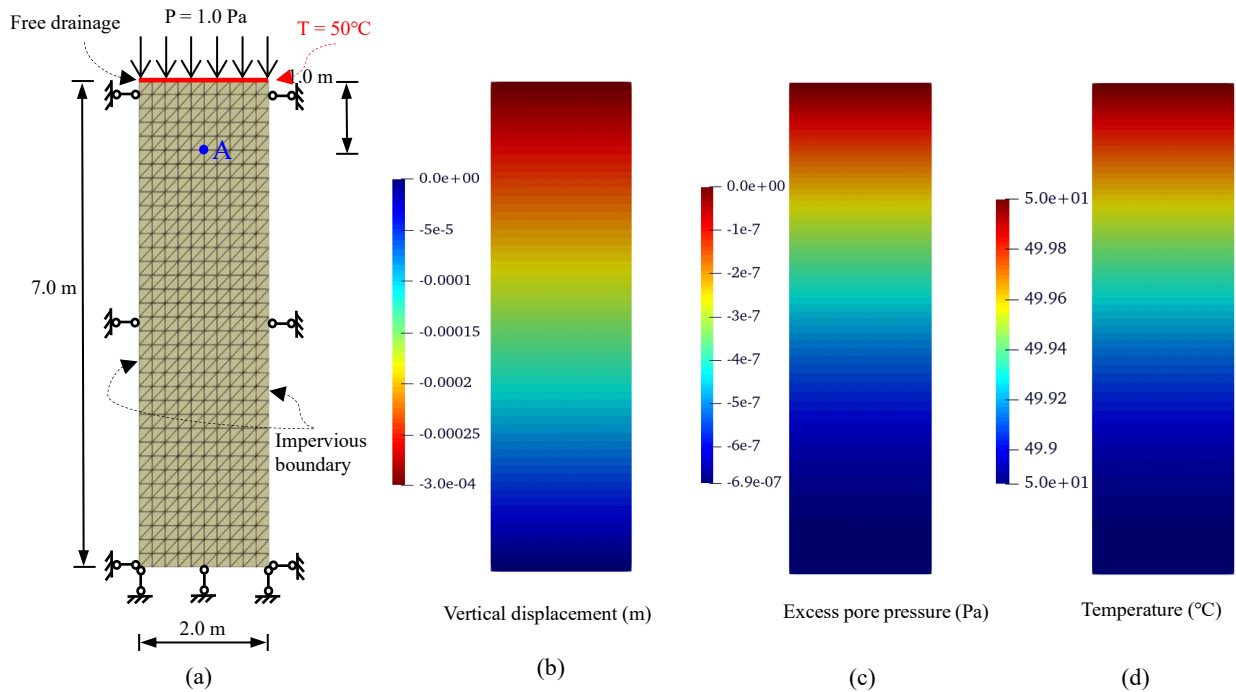

Fig. 13 (a) Thermo-elastic loading model with mesh and boundary conditions, contour of (b)vertical displacement, (c)excess pore pressure, and (d)temperature at the end of consolidation

Fig. 13(b)~(d) illustrated the distributions of vertical displacement, excess pore pressure, and temperature fields at the end of consolidation. At this moment, the vertical displacement field demonstrated a linear distribution, the excess pore pressure had fully dissipated, and the surface temperature had diffused to the bottom of the model. Consequently, the excess pore pressure throughout the model was observed to be near zero, with the temperature uniformly equalized at 50°C.

Fig. 14(a)~(b) depicted the temporal evolution of temperature and excess pore pressure at point A, respectively. The numerical results of bES-FEM showed good agreement with those reported by Cui et al. (2018) and Lewis et al. (1986). A slight difference in the early-stage temperature variations was observed compared to Lewis et al. (1986), which may be attributed to different time marching schemes, as reported by Cui et al. (2018). Fig. 14(c) compared the surface vertical displacement from the bES-FEM with those reported by Cui et al. (2018) and Lewis et al. (1986). Despite minor discrepancies in the peak values, the final displacements after consolidation were nearly identical across these three studies.

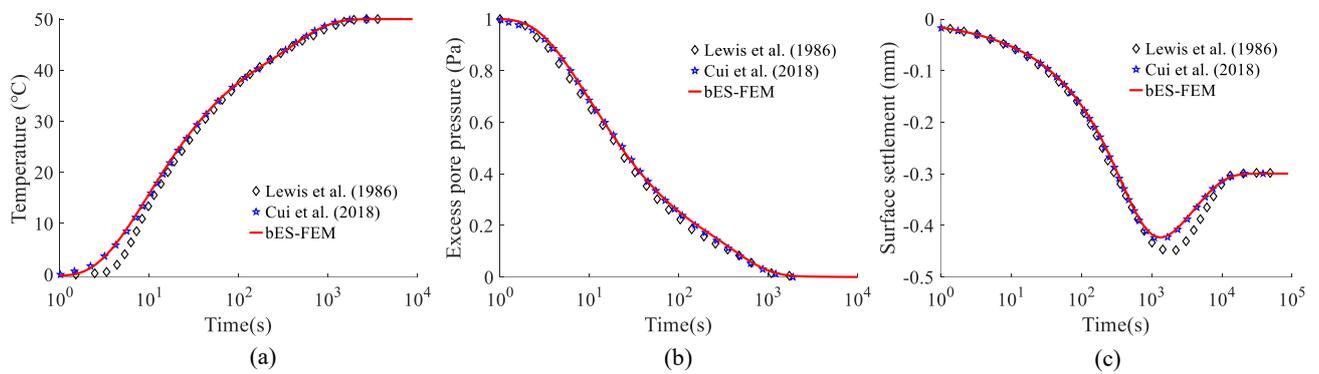

Fig. 14 (a) Variation of temperature, (b) excess pore pressure at Point-A, and (c) vertical displacement at surface

## 5. Applications

Geothermal energy, recognized as a clean and sustainable resource, has attracted growing research attention and seen broad implementation in recent years, particularly in open-loop systems (Antics et al., 2013). These systems extract groundwater for heating or cooling via a heat pump and then re-inject it into the ground at a specified distance. As shown in Fig. 15(a), a dual-well system extracts cold water from an abstraction well for industrial cooling, then returns the warmed water back into an injection well. However, if these two wells are too close to each other, the warm water in the injection well may rapidly convey heat to the cold water in the extraction well, thereby reducing the efficiency of the entire open-loop system. This phenomenon is known as thermal breakthrough (Banks, 2009). The thermal breakthrough time is defined as the moment when the abstraction well temperature is detected to rise by 0.1°C from the initial 10°C. To investigate the determinants of thermal breakthrough in an open-loop well doublet cooling system, a two-dimensional (2D) plane strain coupled model representing a 1m of the 50m aquifer is developed and analyzed using the bES-FEM formulation in this work. The hypothetical scenario explored in this work is informed by Banks (2011), presenting it as a thermo-hydraulic boundary value problem dominated by convection, where the aquifer is assumed to be nearly incompressible and isotropic. Leveraging the model's symmetry, only a half-space representation is modeled, with the finite element mesh illustrated in Fig. 15(b). To mitigate numerical oscillations due to elevated flow velocities and subsequent high Péclet number near the wells, a graded mesh was employed, with the minimum element size of 0.03m around the wells, adhering to the recommendations made by Cui (2015). The abstraction and injection wells maintained a constant flow rate of 0.01m³/s, with an initial temperature set at 10°C and an injection temperature of 20°C. It is assumed that no groundwater or heat flux are in the out-of-plane direction. To reduce the probability of warm water backflow, the injection well is always positioned downstream. The material properties of the aquifer were listed in Table 6.

In this section, we primarily investigated the effects of well separation distance $L$ and hydraulic gradient on thermal breakthrough. For these types of scenarios, Lippmann and Tsang (1980) have suggested an empirical formula to predict the thermal breakthrough time $t_{hdg}$, written as:

if i = 0,

$$t_{hdg} = \frac{\pi L^2}{3Q} \cdot \frac{n\rho_f C_f + (1-n)\rho_s C_s}{\rho_f C_f} \; ; \tag{64}$$

if i ≠ 0,

$$t_{hyd} = \frac{n\rho_f C_f + (1-n)\rho_s C_s}{\rho_f C_f} \cdot \frac{L}{k_f \cdot i} \left[ \frac{\alpha}{\sqrt{\alpha-1}} \tan^{-1}\left(\frac{1}{\sqrt{\alpha-1}}\right) - 1 \right] , \tag{65}$$

where

$$\alpha = \frac{2Q}{\pi \cdot k_f \cdot i \cdot D \cdot L} . \tag{66}$$

In Eqs.(64)~(66), i represents the hydraulic gradient from the abstraction well to injection well, $Q$ is the pumping rate, and $D$ denotes the effective thickness of the aquifer.

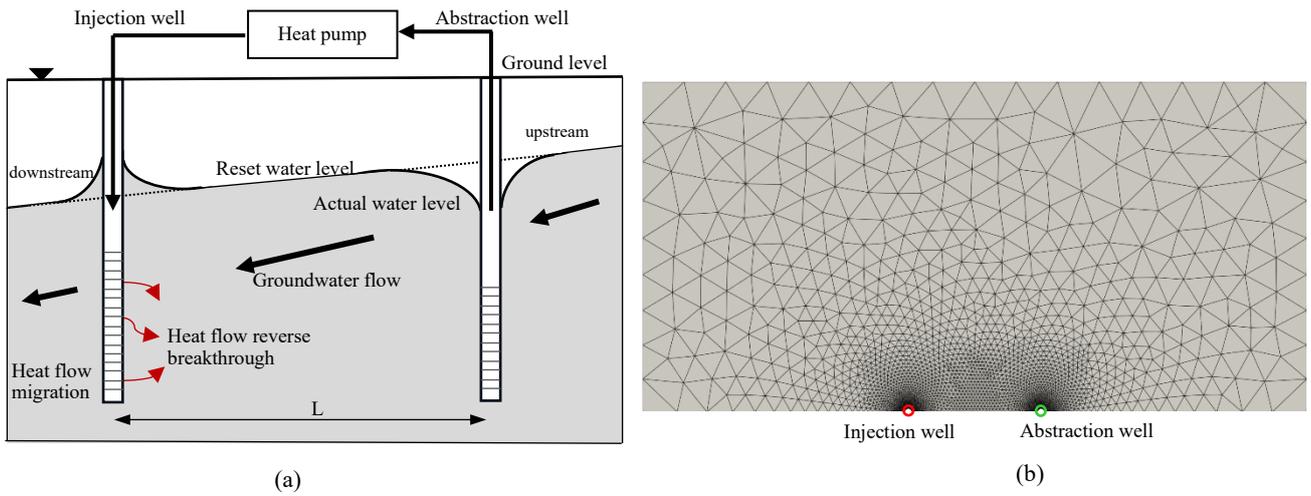

Fig. 15 (a) Open-loop well doublet cooling system; (b) geometric model with mesh

Table 6 Material parameters for open-loop system (Banks, 2011)

| Properties | Symbol | Unit | Value |
| --- | --- | --- | --- |
| Porosity | $n$ | - | 0.091 |
| Specific weight | $\gamma_f$ | kN/m³ | 10.0 |
| Permeability coefficient | $k_f$ | m/s | 5.787e-5 |
| Fluid thermal capacity | $\rho_f c_f$ | kJ/(m³ K) | 2200 |
| Solid thermal capacity | $\rho_s c_s$ | kJ/(m³ K) | 4190 |
| Thermal conductivity coefficient | $k_T$ | W/(m·K) | 1.0 |

The effects of varying hydraulic gradients on temperature breakthroughs were first explored with a well separation distance of 20m. In the initial stage of the analysis, we established a steady state flow between the two wells, and Fig. 16 illustrated the resulting flow pressure fields under hydraulic gradients of 0.0m/m, 0.01 m/m, and 0.02 m/m. Once the steady flow was established, thermal boundaries were then applied to investigate the evolution of the temperature distribution. Fig. 17 exhibited typical temperature contours at a zero hydraulic gradient after 4, 8 and 12 days. The temperature evolution curves at the abstraction well under different hydraulic gradients were depicted in Fig. 18. It can be observed that the case with zero hydraulic gradient exhibited the shortest temperature breakthrough time, indicating that higher hydraulic gradients impede the transfer of heat to the abstraction well, thereby reducing the efficiency of open-loop systems.

The impact of varying well spacings on temperature breakthrough was subsequently investigated. Table 7 presented a comparison between the temperature breakthrough derived from bES-FEM and those predicted by empirical formulas (64)~(66). Despite the temperature breakthrough times from bES-FEM simulations being slightly shorter than those anticipated by empirical formulas, the two were in close agreement. It is notable that increasing the well separation distance can significantly delay the temperature breakthrough, thereby greatly enhancing the system's performance.

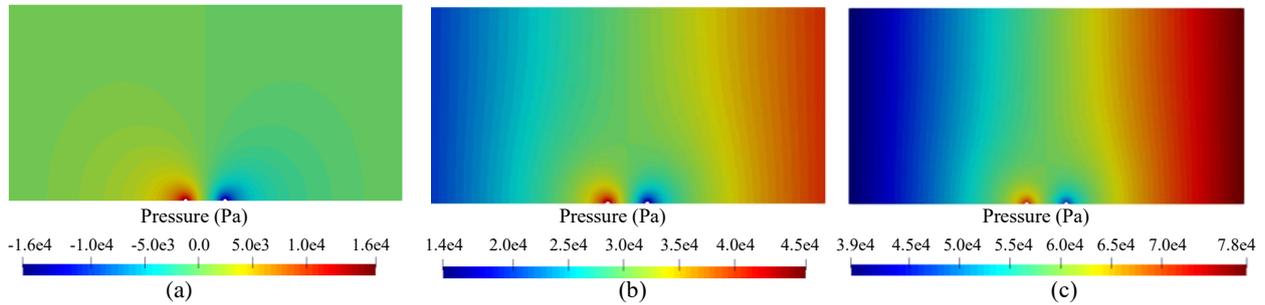

Fig. 16 Pressure field distribution in steady flow with 20m well spacing at hydraulic gradients of (a) $i = 0.0$m/m, (b) $i = 0.01$m/m, and (c) $i = 0.02$m/m (unit: Pa)

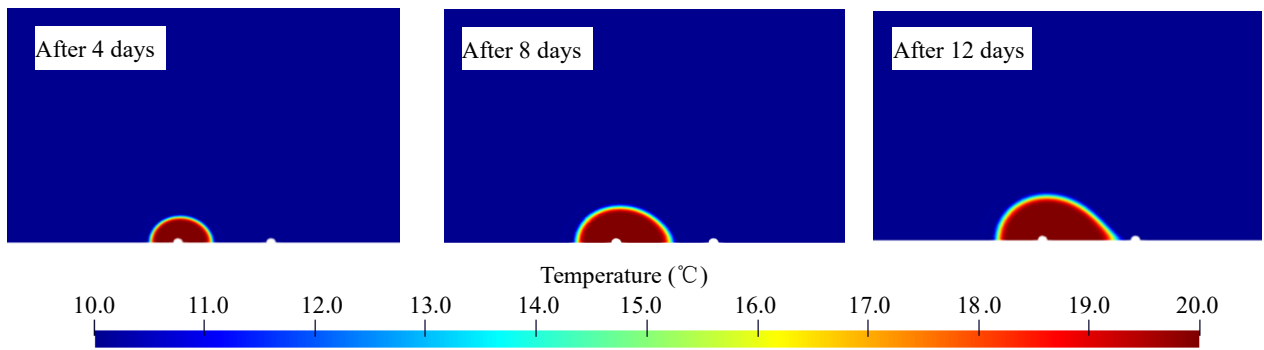

Fig. 17 Temperature field distribution of the aquifer at different time with 0.0 hydraulic gradient and $L$=20m

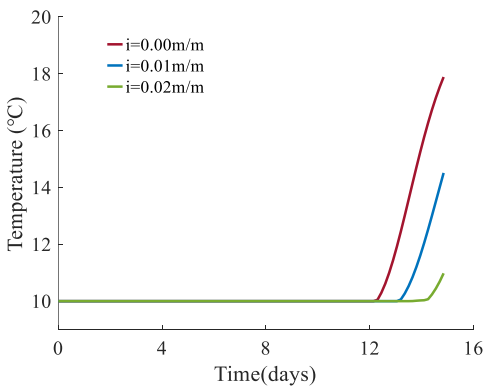

Table 7 Thermal breakthrough time for different analyses

| Well separation distance L(m) | Hydraulic gradient(m/m) | Time by bES-FEM (days) | Time by Lippmann and Tsang (1980) (days) |
|---|---|---|---|
| 20 | 0.0 | 12.35 | 13.77 |
| 20 | 0.01 | 13.45 | 14.86 |
| 20 | 0.02 | 14.76 | 16.15 |
| 40 | 0.0 | 52.78 | 55.10 |
| 40 | 0.01 | 61.95 | 64.60 |
| 40 | 0.02 | 75.42 | 78.54 |
| 60 | 0.0 | 119.63 | 123.97 |
| 60 | 0.01 | 155.11 | 159.38 |
| 60 | 0.02 | 222.64 | 227.97 |

Fig. 18 Comparison of temperature breakthrough at different hydraulic gradients

## 6. Discussions

The current bES-FEM formulation with SUPG stabilization is primarily developed to address two types of numerical instabilities in THM coupled problems: pore pressure oscillations caused by using traditional low-order elements to simulate nearly incompressible, very low-permeability saturated porous media, and thermal instabilities caused by excessive flow velocities in heat transfer processes. However, there is a lack of efforts combining the current methodology to investigate physical instabilities in THM-coupled problems, such as the propagation of cracks inducing pore pressure oscillations during hydraulic fracturing (Cao et al., 2018, 2016; Feng and Gray, 2017; Fisher and Warpinski, 2012; Secchi and Schrefler, 2014). These oscillations primarily stem from intrinsic instabilities caused by physical fracturing, which are different from the numerical instabilities addressed by the current proposed method. Thus, the effect of fracture characteristics on pore pressure oscillation is not considered in this study. Although the proposed method was currently used for solving continuity problems, the bES-FEM formulation can also be used in simulating the simulating crack propagation during hydraulic fracturing process as long as a suitable numerical method for crack simulation is adopted, such as the Finite-Discrete Element Method (FDEM) (Munjiza, 2004). Therefore, the presented method has a wide applicability in practice. Since the authors have deep experience in the development of FDEM (Tang et al., 2024), the proposed THM method can be incorporated into the framework of FDEM to simulate the pore pressure oscillations during the hydraulic fracturing process in the future. Another future work is to apply the proposed bES-FEM method to solve dynamic THM problems.

## 7. Conclusions

An edge-based smooth finite element method with bubble functions (bES-FEM), together with the Streamline Upwind Petrov-Galerkin (SUPG) scheme, has been developed and complemented for coupled THM problems. Compared to using linear triangular elements in conventional FEM, the ES-FEM stands out for its superior performance in terms of high accuracy and convergence rate. By incorporating bubble functions, bES-FEM is further softened and becomes volumetric locking-free for nearly incompressible materials. In addition, the SUPG scheme has been integrated into the current bES-FEM formulation to eliminate spurious oscillations in convection-dominated heat transfer problems.

A progressive series of benchmark tests was conducted to illustrate the effectiveness of the current bES-FEM formulation. The simulation of Terzaghi's 1D consolidation model demonstrated bES-FEM's advantages in simulating nearly undrained limit condition, while the footing consolidation case demonstrated bES-FEM's superiority in mesh independence and its ability to maintain pore pressure stability even with small time steps during simulation. The SUPG scheme's effectiveness in eliminating oscillations was quantitatively validated by comparing results from analytical solutions and numerical simulations with and without SUPG. Two thermo-elastic consolidation problems were finally simulated to illustrate the capacity of bES-FEM in tracking coupled THM problems.

By applying the proposed bES-FEM formulation to open-loop geothermal energy systems, the effects of different well spacings and hydraulic gradients on thermal breakthrough time, which is crucial to the efficiency of the systems, were investigated. A comparison of the numerical results with those obtained from empirical formulas led to the conclusion that bES-FEM can be considered as a powerful and reliable tool for predicting temperature breakthroughs and heat transfer processes in open-loop geothermal energy systems. This was evident from the findings that increasing well spacing and hydraulic gradient in the open-loop well doublet cooling system can effectively delay temperature breakthrough, thereby enhancing the overall system performance.

Future work will focus on improving the bES-FEM formulation with SUPG stabilization in two aspects: (i) extending the current two-dimensional bES-FEM formulation to a three-dimensional version; and (ii) applying the current approach to more complex geotechnical problems involving soil-structure interaction and advanced constitutive models.

## 8. Acknowledgements

This research is financially supported by the National Natural Science Foundation of China (Grant No. 52278363), Shenzhen Science and Technology Program (Grant No.: KQTD20221101093555006), the Research Grants Council (RGC) of Hong Kong Special Administrative Region Government (HKSARG) of China (Grant No.: 15217220, 15229223, N_PolyU534/20) and the Hong Kong Polytechnic University Strategic Importance Fund (ZE2T) and Project of Research Institute of Land and Space (CD78).

**Data Availability Statement`**

All data that support the findings of this study are available from the corresponding author upon reasonable request.

**Conflict of Interest Statement**

We declare that we have no known competing financial interests or personal relationships that could have appeared to influence the work reported in this article.

Zhang, Q., Borja, R.I., 2021. Poroelastic coefficients for anisotropic single and double porosity media. Acta Geotech. 16, 3013–3025. https://doi.org/10.1007/s11440-021-01184-y

Zhang, Q., Wang, Z.-Y., Yin, Z.-Y., Jin, Y.-F., 2022. A novel stabilized NS-FEM formulation for anisotropic double porosity media. Comput. Methods Appl. Mech. Eng. 401, 115666. https://doi.org/10.1016/j.cma.2022.115666